\documentclass[accepted]{article}

\usepackage{microtype}
\usepackage{graphicx}
\usepackage{amsfonts}
\usepackage{amssymb}
\usepackage{amsmath}
\usepackage{subfigure}
\usepackage{amsmath, amssymb}
\usepackage{booktabs} 
\usepackage{xcolor}

\usepackage{hyperref}

\usepackage{icml2020}

\icmltitlerunning{Solving Heterogeneous General Equilibrium Economic Models with Deep Reinforcement Learning}
\newcommand{\expect}{\mathop{{}\mathbb{E}}}
\newcommand{\prob}{\mathop{{}\mathbb{P}}}

\begin{document}

\twocolumn[
	\icmltitle{Solving Heterogeneous General Equilibrium Economic Models \\ with Deep Reinforcement Learning}
			
			
			
	\icmlsetsymbol{equal}{*}
			
	\begin{icmlauthorlist}
		\icmlauthor{Edward Hill}{to}
		\icmlauthor{Marco Bardoscia}{to,ucl}
		\icmlauthor{Arthur Turrell}{to,ons}
	\end{icmlauthorlist}
			
	\icmlaffiliation{to}{Bank of England}
	\icmlaffiliation{ucl}{University College London, Department of Computer Science}
	\icmlaffiliation{ons}{Data Science Campus, Office for National Statistics}
			
	\icmlcorrespondingauthor{Edward Hill}{ed.hill@bankofengland.co.uk}
			
	\icmlkeywords{Machine Learning, ICML}
			
	\vskip 0.3in
]



\printAffiliationsAndNotice{}  

\begin{abstract}

General equilibrium macroeconomic models are a core tool used by policymakers to understand a nation's economy. They represent the economy as a collection of forward-looking actors whose behaviours combine, possibly with stochastic effects, to determine global variables (such as prices) in a dynamic equilibrium. However, standard semi-analytical techniques for solving these models make it difficult to include the important effects of heterogeneous economic actors. The COVID-19 pandemic has further highlighted the importance of heterogeneity, for example in age and sector of employment, in macroeconomic outcomes and the need for models that can more easily incorporate it. We use techniques from reinforcement learning to solve such models incorporating heterogeneous agents in a way that is simple, extensible, and computationally efficient. We demonstrate the method's accuracy and stability on a toy problem for which there is a known analytical solution, its versatility by solving a general equilibrium problem that includes global stochasticity, and its flexibility by solving a combined macroeconomic and epidemiological model to explore the economic and health implications of a pandemic. The latter successfully captures plausible economic behaviours induced by differential health risks by age.
\end{abstract}

\section{Introduction}
One of the core problems in macroeconomics is to create models that capture how the self-interested actions of individuals and firms combine to drive the aggregate behaviour of the economy. These models can provide a guide for policymakers as to what actions they should take in any particular circumstance. Historically, macroeconomic models have tended to be simple because of the need for interpretability, but also because of a heavy reliance on solution methods that are semi-analytical. Such methods allow for the solution of a wide range of important macroeconomic problems. However, events such as the Great Financial Crisis and the COVID-19 crisis have shown that the ability to solve more general problems that include multiple, discrete agents and complex state spaces is desirable. We propose a way to use reinforcement learning to extend the frontier of what is possible in macroeconomic modelling, both in terms of the model assumptions that can be used and the ease with which models can be changed.

Specifically, we show that reinforcement learning can solve the `rational expectations equilibrium' (REE) models that are ubiquitous in macroeconomics where choice variables are continuous and may have time-dependency, and where there are global constraints that bind agents' collective actions. Importantly, we show
how to solve rational expectations equilibrium models with discrete heterogeneous agents (rather than a continuum of agents or a single representative agent).

We apply reinforcement learning to solve three REE models: precautionary saving; the interaction between a pandemic and the macroeconomy (an `epi-macro' model), with stochasticity in health statuses; and a macroeconomic model which has global stochasticity, i.e. where the background is changing in a way that the agents are unable to predict. With these three models, we show that we can capture a macroeconomy that has rational, forward-looking agents, that is dynamic in time, that is stochastic, and that attains `general equilibrium' between the supply and demand of goods or services in different markets.

\section{Background}

Macroeconomic models seek to explain the behaviour of economic variables such as wages, hours worked, prices, investment, interest rates, the consumption of goods and services, and more, depending on the level of complexity. They do this through `microfoundations', that is describing the behaviour of individual agents and deriving the system-wide behaviour based on how those atomic behaviours aggregate. An important class of these models is used to describe how variables co-move in time when supply and demand are balanced (in \textit{general equilibrium}), and when some variables are subject to stochastic noise (aka `shocks'). A typical macroeconomic rational expectations model with general equilibrium is a representation of an economy populated by households, firms, and public institutions (such as the government). The choices made by these distinct agents are framed as a dynamic programming problem in which
households maximise their discounted future utility $U=\mathbb{E}\sum_{t=1}^{\infty}\beta^{t}u\left(s_{t},a_{t}\right)$ with $u$ per-period utility, $\beta$ a discount factor, $s_t\in S$ a vector of state variables, $a_t \in a(s_t)$ a vector of choice variables, and $s_t$ evolving as $s_{t+1} = h(s_t, a_t)$. $\mathbb{E}(\cdot)$ represents an expectation operator, usually assumed to be `rational' in the sense of being the households' best possible forecast given the available information (and implying that any deviations from perfect foresight are random). For household agents, $u$ is monotonically increasing in consumption, $c_t$, and decreasing in hours worked, $n_t$ (both choice variables). Extra conditions are imposed via other equations, for example, a budget constraint of the form $(1+r_t)b_{t-1} + w_t n_t \geq p_t c_t + b_t$ with $p_t$ price, $w_t$ wages, and $r$ the interest rate. $b_t$ captures savings, typically in the form of a risk-free bond or other investment. If $b_t<0$ is permitted (i.e. debt) then $b_t$ usually satisfies a `no Ponzi' condition that rules out unlimited borrowing and effectively imposes the rule that $b_T=0$ ($t\in 0, \dots, T$). Consumers take prices, wages, and interest rate as given; these are state variables. Firms maximise profits $\Pi_t = p_t Y_t - w_t N_t$ (possibly including a $ - r_t K_t$ term if savings are invested) subject to a `production constraint', $Y_t$, that turns labour, $N_t$, and capital, $K_t$ into consumption goods. Typically, $Y_t = A_t f(K_t, N_t)$ where $f$ is a monotonically increasing function of its inputs and $A_t$ is either predetermined or follows a log-autoregressive process $\ln A_t = \rho_A \ln A_{t-1} + \epsilon_t$; $\epsilon_t \thicksim \mathcal{N}(0, \sigma_A)$ is known as a technology `shock'. Governments perform functions such as the collection and redistribution of taxes. Firms are assumed to be perfectly competitive, meaning that each firm takes prices and wages as given.

Prices, wages, and interest rates are determined by \textit{market clearing} for goods, labour, and savings respectively in which supply and demand are balanced in each market. These `general equilibrium' conditions bind agents and the environment together, and are atypical in reinforcement learning. The competitive equilibrium is defined by a vector of state variables, and by consumption and production plans for the agents that maximise utility. Often, the optimal policies of all agents are solved analytically by Lagrangian methods: the equilibrium conditions are substituted in and the system of equations simplified, usually by log-linearising the model around an assumed steady state. We now review some macroeconomic models before briefly discussing multi-agent models more generally.

The \textit{Representative agent with rational expectations} is an important class of macroeconomic model, most well-known is the representative agent dynamic stochastic general equilibrium (DSGE) model. The canonical model is the representative agent New Keynesian (RANK) model \citep{smets2007shocks}. \textit{Continuum rational expectations} models overcome some of the heterogeneity-related shortcomings of those models by replacing the representative household with a continuum of households that are \textit{ex ante} differentiated by their assets and labour productivity. The canonical example is the heterogeneous agent New Keynesian (HANK) model \citep{kaplan2018monetary}. \textit{Macroeconomic agent-based models} differ in that they simulate agents as discrete entities but also typically make very different assumptions to, say, RANK or HANK models; the most important being that they tend \textit{not} to assume rational expectations/perfect foresight and they may not necessarily have competitive markets. Importantly, they allow for heterogeneity in multiple dimensions simultaneously \citep{haldane2019drawing}. \textit{Agent-based models (ABMs)} are also extensively used in epidemiology \citep{abm_epi_rev2018}, sometimes under the name `individual-based models'. At the start of the coronavirus crisis, UK government policy was heavily informed by such models, most notably that of \citet{ferguson2020report}, and there are several ABMs modelling the coronavirus pandemic \citep{hoertel2020stochastic, kerr2020covasim}. These epi-ABMs do not capture economic effects. 

\textit{Epi-macro models} attempt to combine macroeconomic and epidemiological effects, and their interaction. The canonical examples combining epidemiology and a REE representative agent model are \citet{ERT2020} and \citet{eichenbaum2020epidemics} who link the two by assuming that, in addition to the usual Susceptible, Infected, Recovered (SIR) model transmission mechanism as posed by \citet{kermack1927contribution}, a household agent may be infected at work or while engaging in consumption. Market clearing is also assumed. Building on many of the same assumptions as HANK, the canonical continuum agent epi-macro model with REE is by \citet{Moll2020}. Agents are differentiated by their assets, productivity, occupation, and health status. There are three types of good: regular, social, and home-produced; and three types of work: workplace, remote, and home. The epi-macro link is achieved through a transmissibility of infection that is modified to include terms proportional to hours worked and amount consumed, with avoidance of infection captured through a disutility of death. Market clearing is assumed.

Finally, recent work has seen reinforcement learning be applied to multi-agent systems of relevance to economics in the case of bidding in auctions under constraints \citep{feng2018deep,dutting2019optimal}, and deciding on behaviours for both agents and a social planner in a gather-and-build economy \citep{zheng2020ai}.

In the rest of this paper, we show how to use reinforcement learning to solve typical rational expectations macroeconomic models while also incorporating discrete agent heterogeneity and, potentially, stochasticity; demonstrating that all three can be combined is by far our major contribution and has applications for a wide class of economic problems.

\section{Model and Experiments}

\subsection{Precautionary Savings}\label{section_sac}

A typical rational expectations equilibrium problem is that of \textit{precautionary saving}, in which agents anticipate a change in circumstances that will adversely affect their utilities, in this case a reduction in wages, and respond in advance in order to smooth their consumption. Such behaviour is typical of the agents in a REE model. The simplest version of this problem has a known analytical solution. We solve this model using reinforcement learning so that we may compare it to the analytical solution, and we also use it as a way to demonstrate many of the challenges of using RL for this class of problems; notably the speed and accuracy of convergence given the sensitivity to the estimate of the value function; the continuous action and state spaces; and the enforcement of the `no Ponzi' condition. 

\subsubsection{Model}

We assume that there is a single household agent with rational expectations. There are $I=2$ firms, with the firms and the good each firm produces indexed by $i$. The household agent is employed by one of these firms, which we will denote $e$, and has per period utility $u_t = \sum_{i \in I}\ln c_{it} - \frac{\theta}{2}n_t^2\label{util-pc}$ with action (choice variables) $c_{it}$ consumption and $n_t$ hours worked. $0 \leq t < T$ is the discrete timestep. The price vector is fixed to $p_{it} =1$, and the interest rate is fixed to $0$. The wage is imposed as $w = 1$ for $t<T/2$ and $w = 0.5$ afterwards, a fall that is anticipated. The household agent is subject to a budget constraint such that $b_{t+1} = b_{t} + w_t n_t - \sum_{i \in I}p_{it}c_{it}$. The no Ponzi condition is imposed via $ b_T = 0$, which prevents unlimited borrowing by the household. The agent maximises its discounted utility $\sum_{0\leq t<T} \beta^t u_t$ with $\beta = 0.97$ and $T=20$.

We use $I=2$ firms rather than a single representative firm, allow $\theta$ (the utility of work) to vary, and introduce four discrete states, indexed by $d$, that the household can be in at any given time. While making no economic difference, the expansion of the action and state spaces means that the hyperparameters we find are transferable to the more realistic problems we will come to. It also allows us to test that training the parameters of a single network to provide the value function for any agent is a good approximation to training a network for each agent individually, which is significantly more computationally expensive.

The analytical solution for the consumption and hours paths for the household are given by $c_t = \lambda^{-1}\beta^t / I$ and $n_t = w_t \theta^{-1}\lambda \beta^{-t}$ where $\lambda$ is a Lagrange multiplier determined from the no Ponzi condition. Computationally, we start from the Bellman equation:
\begin{multline}
   U(t,s,S) = \\ \max_a \left\{u_t(s, S, a) + \beta \displaystyle\expect_{s_{t+1},S_{t+1}|a,s,S}U(t+1,s_{t+1},S_{t+1})\right\}\label{bell-base}
\end{multline}
where $s$ is the agent's state, $a$ is the action vector. $\mathbb{E}$ is the expectation operator, and $S$ is the global state (which includes $t$, but we make $t$ explicit for clarity). For the current problem, we drop the global state to obtain $U(t,s) = \max_a \left\{u_t(s, a) + \beta \expect_{s_{t+1}|a,s}U(t+1,s_{t+1})\right\}$.

The optimal action vector under local and global constraints, $a_t^*(s, S)$, is computed using the method of Lagrange multipliers; this requires accurate values of the gradient of $U$ with respect to the state variables.

We use a deep neural network to approximate $U(t, s) = U(t, s= (e, \theta, d; b_t))$; however, direct approximation is problematic because $\partial_{s} U(t, s)$ is slow to converge and is highly sensitive to initialisation and hyperparameter choice. To mitigate this, we find $D(t, s) = \partial_{s} U(t, s)$ explicitly by solving
\begin{multline}
	D(t, s) = \partial_{s} u_t(s, a^*) \\+ \beta\sum_{s_{t+1}} \partial_{s} \prob(a^*(s)\rightarrow s_{t+1}|s,a^*) U(t+1,s_{t+1}) \\+ \beta\expect_{s_{t+1} |s,a^*} \partial_{s}(s_{t+1}) D(t+1,s_{t+1})\label{bell-d}
\end{multline}
That a directly learnt estimate of the $\partial_{s} U(t, s)$ aids stability and convergence has been noted in both deterministic \citep{balduzzi2015compatible} and stochastic \citep{heess2015learning} continuous control problems.

In the case of precautionary saving, $\partial_{s} \prob(s\rightarrow s_{t+1}|s,a^*) = 0$ and $\prob(s\rightarrow s_{t+1}|s,a^*) = \delta_{ss_{t+1}}$, (with $\delta$ the Kronecker delta) so that $D(t, s) = \partial_{s} u_t(s,a^*) + \beta\frac{\partial s_{t+1}}{\partial {s}} D(t+1, a^*(s))$. The values of $\partial_{a} U(t+1, s_{t+1})$ that are required for finding $a^*$ are then written as $\frac{\partial {s_{t+1}}}{\partial {a}} D(t+1, s_{t+1})$. $U$ and $D$ need not be consistent.

\subsubsection{Specification}\label{spec-base}

\emph{System} All timings use a laptop with a 4-core CPU (an i7-6700HQ) without GPU acceleration. The code is written in \textsc{Python3} using \textsc{PyTorch} \citep{NEURIPS2019_9015} for the neural network. 

\emph{Neural Network and Training} The networks for $U$ and $D$ are identical with $5$ layers of $50$ Softsign neurons, followed by a single linear layer. The inputs are normalised to the typical scales in the problem, for example $t \rightarrow (2t - 1) / 2T$, and the outputs of each network are normalised to the scale of the problem by an additive and multiplicative factor. 
The networks are wrapped in a caching and linear interpolation function to reduce network evaluations.

The networks are trained using the Adam optimiser \citep{kingma2014adam} with a decaying learning rate over epochs, $E$ of $l_r = \max(5 \times 10^{-3} \times 0.8^{-E}, 10^{-5})$. Each epoch contains $160$ experiences of $U(t, s)$ and $D(t, s)$, which are trained with a replay buffer \citep{mnih2015human}. Initially the buffer is emptied after each epoch, but once $l_r \leq 10^{-4}$ it retains a fraction of its contents. This provides swift initial learning followed by good coverage of experiences later in the process \citep{fedus2020revisiting}.

The experiences are created using $n=4$-step learning \citep[e.g.][]{sutton2018reinforcement}, recorded by running an agent forward taking its current optimal actions. These are run independently, in parallel. $n/(n+1)$ of forward runs apply a perturbation to the state every $n$ steps, allowing for exploitation and exploration. In early epochs, where the predicted solution is less accurate, Double DQN \citep{van2016deep} and target clipping \citep{schulman2017proximal} provide stability. 

\subsubsection{Results}
We assess the goodness of the model by defining its error as the mean absolute fractional difference between the analytic consumption from the equations above and the simulated consumption for an agent beginning at $t=0$ with $b_0 = 0$ for a number of values of $\theta$. This is a stringent and appropriate test of the model since the value of consumption at each timestep depends on current savings, which are themselves determined by the time-histories of $n$ and $c_i$. Note that this means errors in the observable values compound over time, as they will in similar models in later sections. The model successfully converges to the analytical solution with error of $\approx 0.01$ in $\approx 25$ epochs, with each epoch containing $160$ experiences. This takes $\approx 2$ minutes, about $1$ GFLOPs-hour.

\subsection{A General Equilibrium Rational Expectations Epi-Macro Model with Stochasticity in Agents' States}\label{section_epi}

We now build on the previous example to demonstrate the solution of a rational expectations general equilibrium \textit{epi}-macro model, with SIRD health states. We find the `decentralised equilibrium' in which each household agent behaves optimally according to its choice variables.
Agents are motivated to change their behaviour due to fear of dying from the disease and, because consumption carries with it a risk of contracting the disease, their patterns of consumption change, in turn altering their risk of infection -- this risk therefore connects the macroeconomic and epidemiological aspects of the model. These consumption changes are differentiated by age as there is an exponentially increasing risk of death according to age once infected.

\subsubsection{The model}
The model combines features of both agent-based macro models and rational expectations models. Agents are rational, forward-looking, and discrete. Let household agents be indexed by $j$, while sectors (the analogue of firms), and the goods that each sector produces, are indexed by $i$. Household are \textit{ex ante} differentiated by their age and the sector that employs them. Let $E_i$ be the set of households employed by sector $i$. 

The model is a real business cycle (RBC) model \citep{kydland1982time}, with no technology shock. In each period household agents engage in consumption, $c$, and work, captured by hours worked, $n$. Time-$t$ utility of household $j$ when susceptible, infected, or recovered is $u_j = \sum_i \ln c_{ji} - \frac{1}{2}\theta n_{j}^2$ where $\theta=1$ is a disutility of working. Households face a time-$t$ budget constraint balancing their income from work and interest on the capital they have loaned to industry, against their consumption and investment in new capital, $v_j$: $w_j n_j + rk_{j,t} = \sum_i p_i c_{ji} + v_{j,t}$, with $k_{j,t+1} = k_{j,t} + v_{j,t}$. Sector $i$ produces a quantity of goods $Y_i$ using household labour such that $Y_i = A N_i^\alpha K_i^{1-\alpha}$ where $N_i$ are the total hours worked in sector $i$ and $K_i = K_e + \hat{K}_i$ is its total capital. $K_e$ is an initial endowment of capital and $\hat{K}_i$ is that derived from consumers' investments.  We set $A=1$, $\alpha=2/3$. Sectors are profit-maximising so $\partial_{N_i}\Pi_i = 0$ and $\partial_{K_i}\Pi_i = 0$ where the profit $\Pi_i = p_i Y_i - w_i N_i - (r + \delta)K_i$, with $\delta$ the depreciation rate of capital. For our form of $Y_i$ this implies $\Pi_i = 0$. All consumers and firms take $p_i$, $w_i$ and $r$ as given, and these are adjusted to clear the markets for hours worked: $\sum_{j \in E_i} g_j n_j = N_i \; \forall i$ and capital: $\sum_j g_j k_j = \sum_i \hat{K}_i$, where agent weights $g_j = J^{-1} \;\forall j$. The utility of death is $-200$, and the discount rate is $\beta = 0.97$. See Appendix I for a more detailed description. 

This type of model, which is common, is used only to demonstrate the approach; the solution method is applicable to a wide range of models. Also, note that there is no need for linearisation around a steady-state, which is a common solution technique for models of this type.

Household agents have four possible health states: susceptible, infected, recovered, and deceased (SIRD), with the probability of transition between states given by, e.g., $\mathbb{P}(S\rightarrow I)$ for going from susceptible to infected. In what follows, epidemiological parameters are noted with tildes. At each time-step (a day),
\[ 
    \mathbb{P}_j(S\rightarrow I) = \tilde{\beta} \sum_i\frac{\tilde{\rho}_i c_{ji}}{c_{ji,t=0}}\frac{\displaystyle\sum_{\text{Infected }j} c_{ji}}{\sum_j c_{ji}}
\]
so that there is a `shopping risk' of acquiring the infection if many infected are consuming the same goods. $\tilde{\rho}_i$ is a vector of relative consumption risks such that $\sum_i \tilde{\rho}_i = 1$. $\mathbb{P}(I\rightarrow R) = \tilde{\gamma}$, and $\mathbb{P}_j(I\rightarrow D) = \tilde{\xi}(j)$ for each household $j$ where $\tilde{\xi}(j)$ is an exponentially increasing function of age rising from $0.006$ at age $40$ through $0.024$ at $55$ to $0.165$ for a 70 year old, then flattening at age $80$. $\tilde{\beta}=0.56$, and $\tilde{\gamma}=0.2$ from \citet{lin2020conceptual}. The relative risk of consuming each sector's product is $\tilde{\rho}_i = \{0.8, 1.2\} / 2$, and we refer to these as `remote' and `social' consumption respectively. At time $t=0$, $k_j = 15 = K_e / 2  \;\forall j$, and at time $t=2$ we infect the youngest $10\%$ of the population. On death, any investments are redistributed across living agents and deceased agents are not replaced. We use $J=100$ agents, with a distribution of ages given by $\text{Age}(j) \thicksim \mathcal{U}(20, 95)$. Households are evenly distributed across employers, and cannot change employer.

Aside from using a reasonable distribution of the death rates, this model is entirely uncalibrated.

\subsubsection{Solving the model with multiple agents}

We run a number of simulations indexed by $\tau$. Solving the model means finding a history $H_{\tau = \infty} = \{S_t\}_{t \in 0, ..., T}$ and agent behaviours $U_{\tau = \infty}(t,k_t,S_t)$ that are consistent.  We begin with $H_{\tau = 0}$ and $U_{\tau =0}$. We use $\bar{H}_\tau$, an average created from $\{H_{\tau'}\}_{\tau' \leq \tau}$ to re-train $U_\tau$, obtaining $U_{\tau+1}$. A multi-agent simulation is run with agent behaviours governed by $U_{\tau+1}$ to obtain $H_{\tau+1}$ in general equilibrium. As is standard in iterative methods, we terminate at a sufficiently large $\tau_{max}$ in order to provide a good approximation to the values at $\tau = \infty$; the results we present use $\tau_{max} = 50$. In the limit of large $J$, each household's behaviour makes no difference to the system, so despite the stochastic transitions of health statuses, we are able to obtain a unique history of the variables characterising the global system (including the $I=2$ sectors). The observed $\hat{H}_{\tau}$ can be seen as noisy observations of that $H_{\tau}$, and $\bar{\hat{H}}_\tau$ as an estimate of $\bar{H}_\tau$ where the averaging is chosen such that $\bar{H}_\tau \xrightarrow{\tau\rightarrow \infty} H_{\tau}$ and $\bar{\hat{H}}_\tau$ has significantly less noise than $\hat{H}_\tau$. We use the average of $\{H_{\tau'}\}_{\tau'\leq \tau}$ weighted by $\gamma^{\tau -  \tau'}$ with $\gamma \leq 1$. While in general $H_\tau$ is therefore produced solely as a function of $U_{\tau-1}$, we modify this general scheme for this specific case by providing the infection fraction $\sum_{\text{Infected }j} c_{ji} / \sum_j c_{ji}$ in multi-agent simulation $\tau$ from $\bar{H}_{\tau-1}$. This counteracts the propagation of the high levels of noise created by the initiation of the pandemic through to later times of the simulation, and could be avoided by using a larger number of agents.

At each timestep, we iteratively solve to find the values of the prices $(p_i, w_i, r)$ for which the markets for labour and capital clear by gradient descent using \texttt{scipy.optimize.least\_squares} with the default parameters, initialised from the previous timestep, for $t>0$. We then advance both the capital and epidemiological state to proceed to the next timestep.

\subsubsection{Re-training agent behaviours}

$U_{\tau+1}$ is obtained from $U_{\tau}$ by continuing training using $\bar{H}_\tau$. The network parameters are the same as in \S \ref{section_sac}, however we use a gentler decay of the learning rate. There are no problems observed with convergence, and the adherence to the no Ponzi condition is a good test of this, since achieving it is sensitive to the entire time history of the simulation.

As in the multi-agent model, consumers take prices ($p_i$,$w_i$ and $r$) as given, find their optimal consumptions, hours worked and investments before advancing their state using the budget constraint and the probabilities of their SIRD state changing.

\subsubsection{Results}

We examine two cases: a `heterogeneous' case as described above, and a `homogeneous' case without age heterogeneity but with the same mean death rate. 

\begin{figure}[ht]
\vskip 0.2in
\begin{center}
\centerline{\includegraphics[width=\columnwidth]{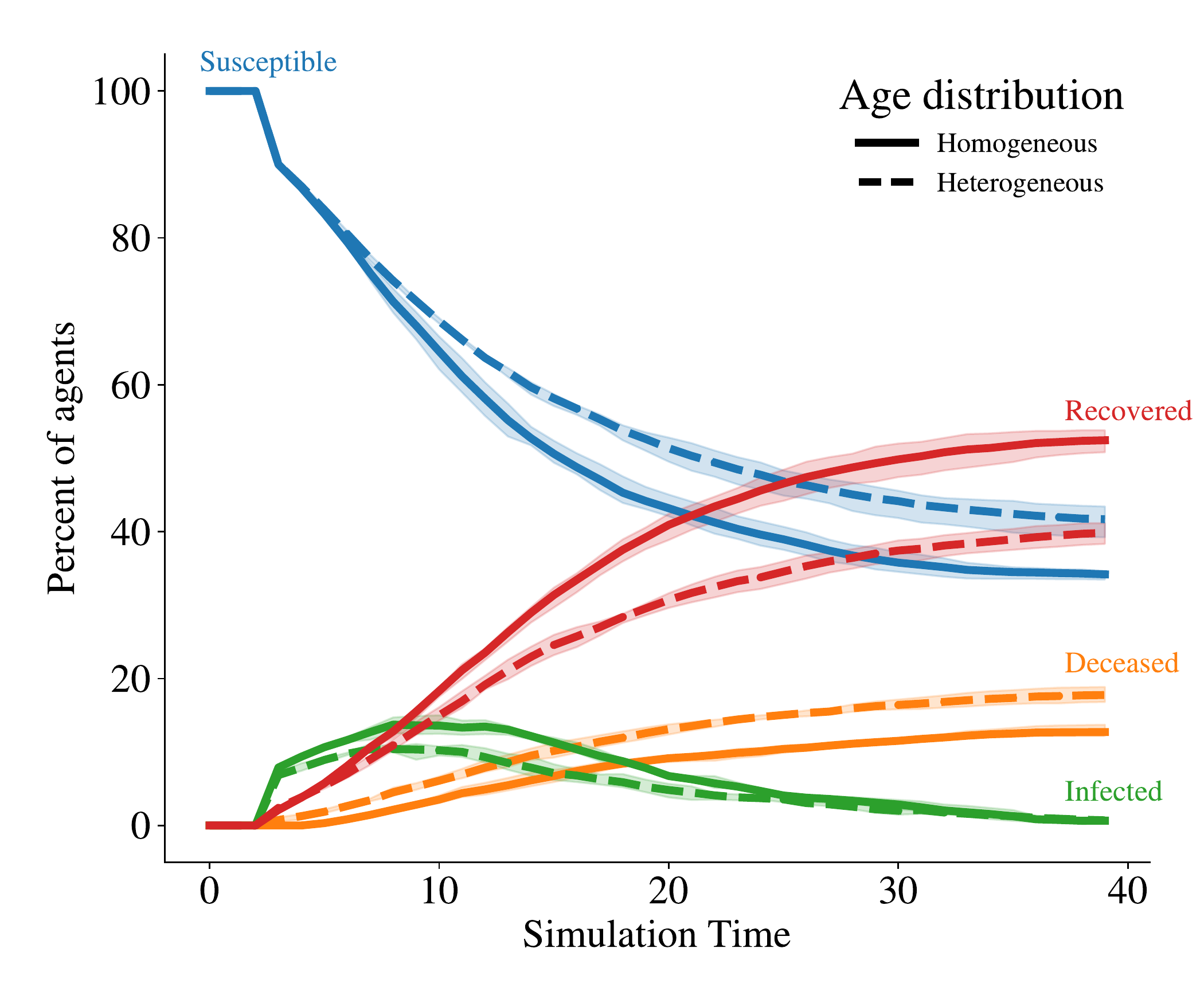}}
\caption{Percentage of agents who are susceptible, infected, recovered, or deceased as the simulation progresses.}
\label{fig-sird}
\end{center}
\vskip -0.2in
\end{figure}

Figure \ref{fig-sird} shows the percentages of susceptible, infected, recovered and deceased agents as the pandemic progresses. Each line is an average over the results of $3$ simulations, and each simulation's result is an average over $20$ histories. As can be seen from the 95\% confidence intervals in the figure, the behaviour is similar across simulations. In the homogeneous-age case, more people are infected over the course of the pandemic, but there are fewer deaths in total.

\begin{figure}[ht]
\vskip 0.2in
\begin{center}
\centerline{\includegraphics[width=1.1\columnwidth]{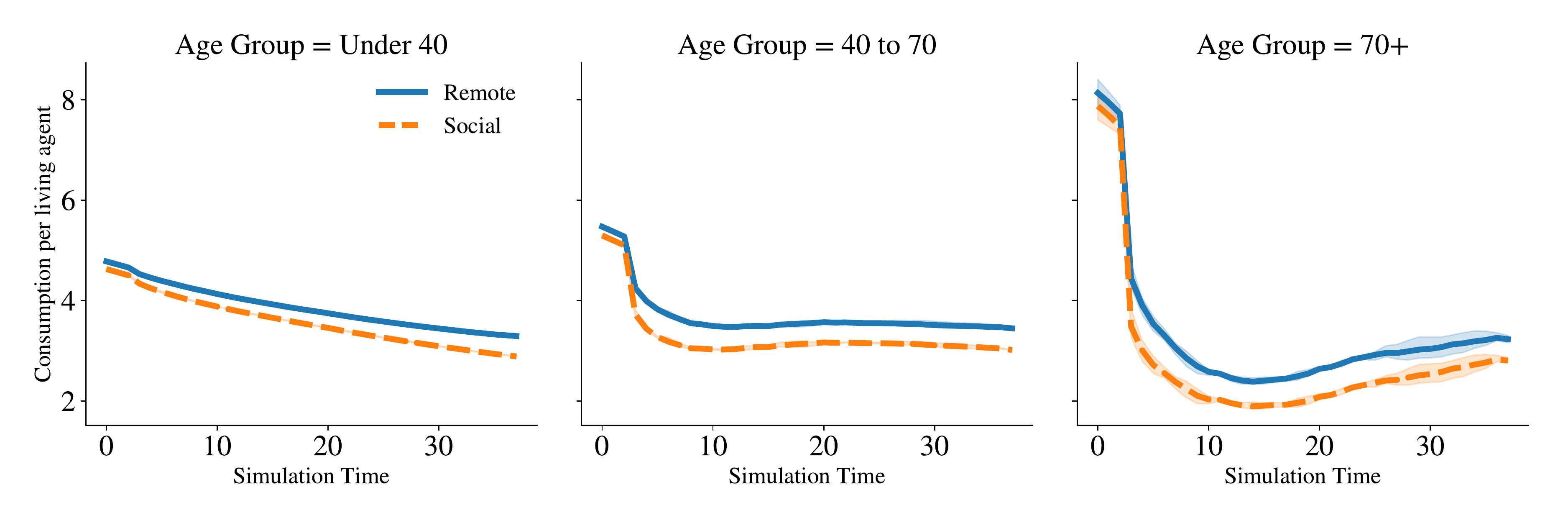}}
\caption{Average consumption of living agents from the `remote' and `social' sectors binned into three age groups.}
\label{fig-consump-age}
\end{center}
\vskip -0.2in
\end{figure}

Figure \ref{fig-consump-age} shows the agents' consumptions. We bin the uniform age distribution into young ($<40$), old ($>70$) and middle-age groups; we find considerable differences in behaviour between them. After consuming the most before the pandemic since they anticipate the coming opportunity to save, the old strongly reduce consumption in response to infection risk, and reduce consumption of the riskier `social' good more. The young, conversely, are unlikely to die and so their consumption is relatively unchanged, governed by the decrease in the size of the economy. 

\begin{figure}[ht]
\vskip 0.2in
\begin{center}
\centerline{\includegraphics[width=\columnwidth]{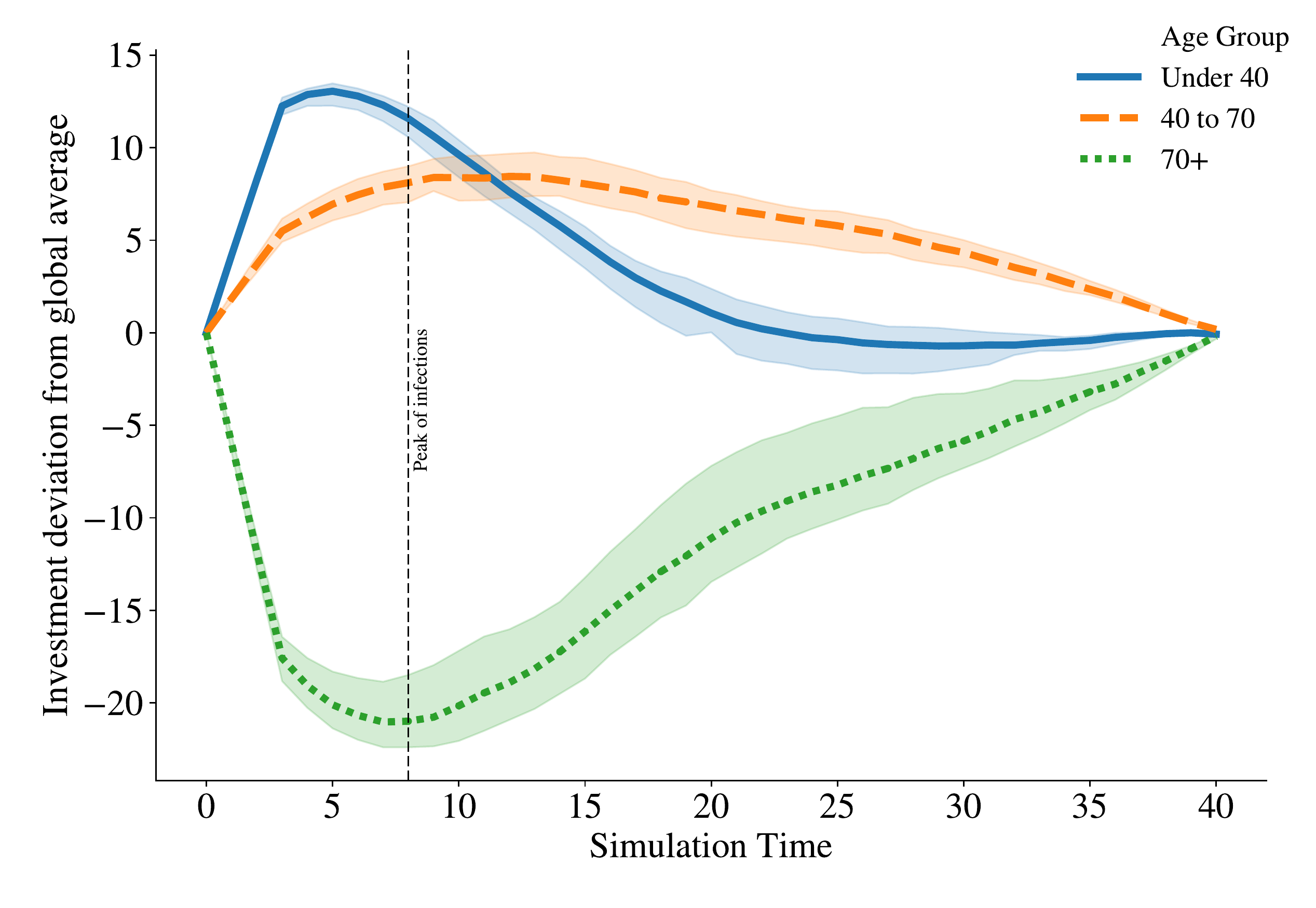}}
\caption{Average investments of living agents binned into three age groups. The vertical line shows the time of peak infections.}
\label{fig-savings-age}
\end{center}
\vskip -0.2in
\end{figure}

Figure \ref{fig-savings-age} shows the mean investments per agent in each age group, normalised to the salary (product of wage and hours worked) of an agent in the same system with no pandemic and no saving. The young anticipate the pandemic and save before it in order to spend when infection rates are higher, with the converse behaviour for the old. It also shows the no Ponzi condition holds with good accuracy for all age groups.

\begin{figure}[ht]
\vskip 0.2in
\begin{center}
\centerline{\includegraphics[width=\columnwidth]{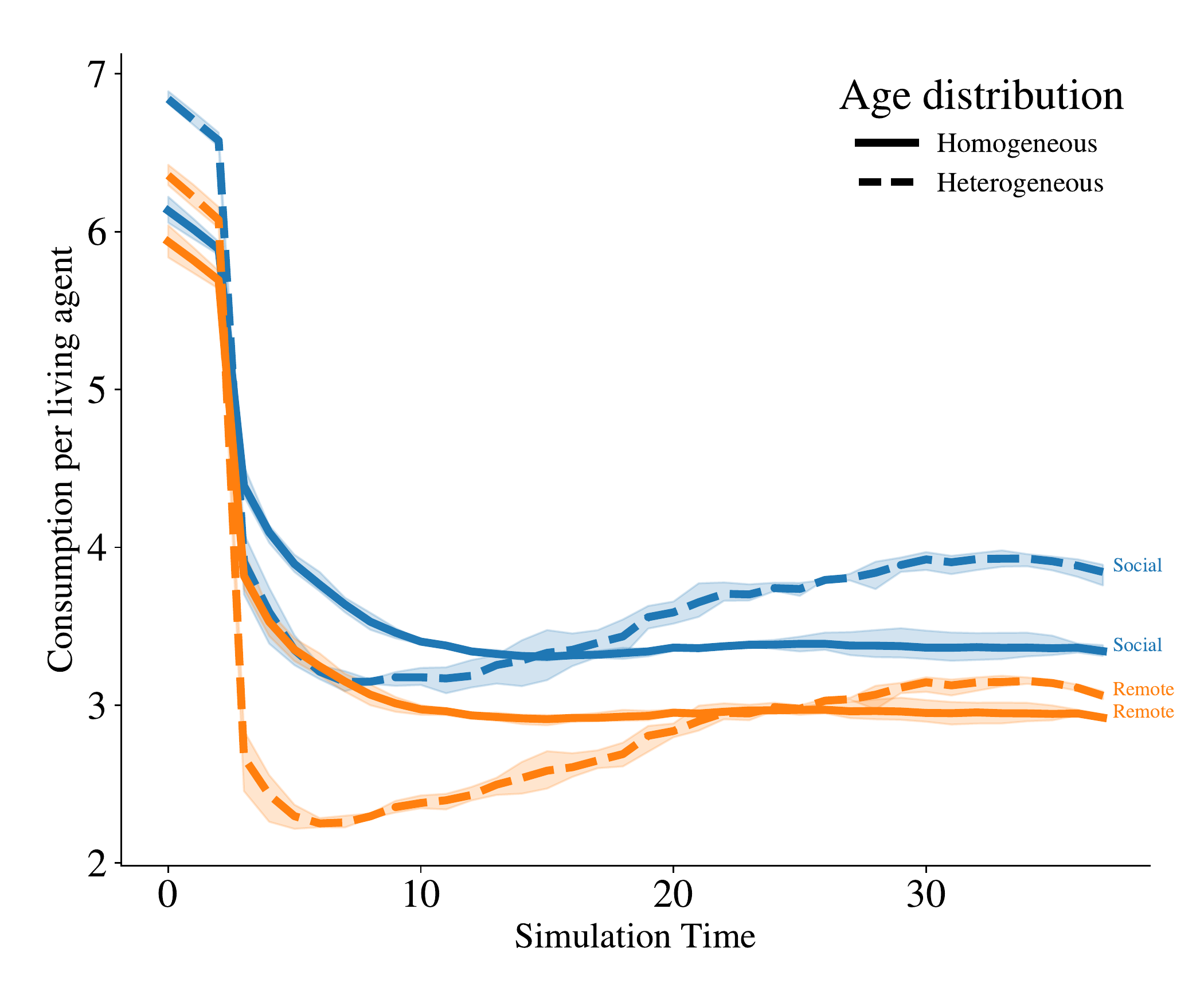}}
\caption{A comparison of total consumption per living agent by sector in two separate simulations with agents who are heterogeneous or homogeneous in age.}
\label{fig-cons_h_h}
\end{center}
\vskip -0.2in
\end{figure}

Finally, Figure \ref{fig-cons_h_h} shows the total consumption for both heterogeneous-age and homogeneous-age cases. The inclusion of distributional effects causes significant changes to bulk macroeconomic quantities. 

Together these results show that in this uncalibrated model the inclusion of age heterogeneity makes a substantial difference to both the epidemiological and economic progress of a pandemic. This model and solution method has been tested with a range of epidemiological and economic parameters and has shown consistent stability and convergence. This exercise has also shown the sensitivity of the model's conclusions to those parameters, emphasising the importance of calibration in all aspects of the model if it were used as more than a test case of the methodology.  

\subsubsection{Specification}

The hardware, software, and parameters are identical to \S\ref{section_sac} with the exception of the learning rate decay which is slower here to allow for the longer time history.  Scaling is linear with $J$, since the number of iterations of the least squares optimisation seems to scale very weakly with $J$ for this problem. Each history calculation followed by an RL update takes $\sim 30$ minutes on the reference machine, so a single simulation takes $\sim 24$ hours, $\sim 720$ GFLOPs-hour.

\subsection{Stochasticity of global variables: A toy wage-shock problem}\label{section_global}

In the previous section, only the agents' state was stochastic and, because of being in the limiting case of large numbers of agents, each individual agent's state did not affect the global state. We now return to the model in \S \ref{section_sac}, but instead of having a deterministic drop in wages, wages now follow a log-autoregressive stochastic process.

\subsubsection{The model}

We return to the Bellman equation, \eqref{bell-base}. We assume there is no stochasticity in health states so the $\expect_{s'|s,a}(\cdot)$ are no longer present, however the $\expect_{S_{t+1}|S}(\cdot)$ remain.

\begin{multline}
	U(t,s,S) = \\ \max_a \left\{u_t(s, S, a) + \beta \expect_{S_{t+1}|S}U(t+1,s_{t+1},S_{t+1})\right\}
\end{multline}
where $s_{t+1} = a(s)$ advances deterministically. We change the method to solve for $\tilde{U}$ and $\tilde{D}$, defining 
\begin{equation}
\tilde{U}(t,S,s_{t+1}) = \beta \expect_{S_{t+1}|S} U(t+1, S_{t+1},s_{t+1})
\end{equation}
and
\begin{equation}
U(t,S,s) = \max_{a}\left\{ u_t(S,s,a) + \beta\tilde{U}(t, S,s_{t+1})\right\}
\end{equation}
where the value of $a$ which attains the maximum defines $a^*$. Again, we will find the maximum using Lagrange's method and the auxiliary quantity $\tilde{D}$,
\begin{equation}
\tilde{D}(t, S,s_{t+1}) = \expect_{S_{t+1}|S} \left[D(t+1,S_{t+1},s_{t+1}) \right]
\end{equation}
and so
\begin{multline}
    \partial_a \mathcal{L}(t, S, s, a) = \partial_a u_t(S, s, a)\\ + \beta (\partial_a s_{t+1}) \tilde{D}(t+1, S, s_{t+1})
\end{multline}

As is standard in reinforcement learning, the expectations are approximated by using a large number of global state histories to update $\tilde{U}$ and $\tilde{D}$.

Excepting the wage history, the set-up is as in \S \ref{section_sac}. The agents are statically heterogeneous in their propensity to work, $\theta \in [0.6,1.4]$, and employer, $e$; and dynamically heterogeneous in savings. Again, the multiple employers and the treatment of heterogeneity in $\theta$ are introduced so that demonstrations of convergence and hyperparameter choice are relevant to the problem in the next section.

In this toy problem, we compare two types of agent: one a current-time \emph{wage-observing} agent, whose future utility is a function of the wages at the current timestep, $w_t$: $U(t, S = (t,w_t), s = (\theta, e, b_t))$; the other a \emph{non-wage-observing} agent with $U(t, S = (t), s = (\theta, e, b_t))$. Both are able to optimise by testing how their strategies play out within the histories they have seen, however they only have partial visibility of the global state. In the previous case that had deterministic global state, a knowledge of the time determined all other global variables, but here the mapping from observable values to global state is one-to-many. 

The wage follows a log-autoregressive process,
$\ln w_t = \rho_w \ln w_{t-1} + \epsilon_t$; $\epsilon_t \thicksim \mathcal{N}(0, \sigma_w)$; where we use $\rho_w = 0.97$, $\sigma_w = 0.1$, and $w_{t=0} = 1$. Since the wage is autoregressive, knowledge of the current wage adds information about the wage in the future. The agent is trained on $\# T \in \{100,1000\}$ training histories.

We parametrise wage histories by their mean absolute fractional deviation, $d_h$, from the mean path of the auto-regressive process $w_{\text{mean},t} = \exp \left(\frac{\sigma^2_w}{2} \sum_{t' < t} \rho_w^{2t'}\right)$. Of the $50$ test histories, which have $d_h \in [0.07, 0.55]$, we consider the $27$ with $d_h > 0.2$ to represent those with significant deviation from the mean path.

We judge the success of this model by calculating the average total utility an agent with $\theta = 1$ attains over these previously unseen wage histories, $\{w_{h,t}\}_{0 \leq t < T}$, where the average over histories with $d_h > x$ is denoted $\bar{u}_{d_h > x}$. Table \ref{tab-wage} compares the average utilities of agents with different training setups and wage visibility to an analytic approximation found by defining the action at time $t$ in history $h$ to be the values obtained from the formulae in \S \ref{section_sac} for a wage history beginning at time $t$: $c_{h,t} = 1/\bar{\lambda}$ and $n_{h,t} = w_{h,t} \theta^{-1}\bar{\lambda}$, obtaining $\bar{\lambda}$ from the no Ponzi condition as $I^{-1}\sum_{t' \geq t} \beta^{t' - t}  = \bar{\lambda}^2\theta^{-1} \sum_{t' \geq t}\left\{\beta^{t-t'}\expect_p w_{p,t'}^2 \right\}$. $\expect_p w_{p, t'}^2$ is evaluated analytically over all possible wage paths that have $w_{p,t} = w_{h,t}$ using the second moment of the log-normal distribution. 

\begin{table}[t]
\caption{Total utilities achieved by different agents relative to the analytic approximation for different visibility of the wage, numbers of training histories $\#T$ and including prioritised experience replay (\textsc{per}). }
\label{tab-wage}
\vskip 0.15in
\begin{center}
\begin{small}
\begin{tabular}{lccc}
\toprule
Agent & $\#T$ & $\bar{u}_{d_h\geq 0}$ & $\bar{u}_{d_h > 0.2}$  \\
\midrule
Wage-observing + \textsc{per}& 1000  & \textbf{+0.97} & \textbf{+1.21} \\
Wage-observing & 1000 & +0.54	& +0.29\\
Wage-observing + \textsc{per}& 100 &  +0.27 & +0.17 \\
Analytical approximation & - & 0.0 & 0.0 \\
Non-wage-observing + \textsc{per} & 1000 & -0.39 & -1.43 \\
\bottomrule
\end{tabular}
\end{small}
\end{center}
\vskip -0.1in
\end{table}

We implement prioritised experience replay \citep{schaul2015prioritized} by retaining experiences with a larger error for a larger number of training epochs. This increases the agents' performance relative to a base agent trained as in previous sections, particularly on the $d_h > 0.2$ histories that deviate significantly from the mean path. This is expected since the training examples are sparser and more varied at higher $d_h$. The wage-observing agent outperforms the non-wage-observing agent. In addition to the solution becoming stable and the no Ponzi condition being satisfied, that the utilities for the wage-observing agent are consistently higher than those for the analytical approximation gives us confidence that the answers converged to are accurate. We record the utilities after $100$ epochs which equates to $40,000$ experiences or $10$ minutes on the reference machine. A degradation in performance is seen if an insufficient number of training histories are used.

\subsubsection{Specification}

The specification is identical to \S \ref{spec-base} except that the rate of decay of the learning rate is decreased to allow averaging, $l_r = e^{-0.01E} / (1+E)$  for epoch $E$; and, as discussed, prioritised experience replay is used. 

\subsection{Stochasticity of global variables: A general equilibrium with stochastic technology shocks}\label{section_dsge}

Finally we demonstrate a general equilibrium model that has stochastic global variables, specifically a log-autoregressive process in the technology. This is given by $\ln A_{it} = \rho_{A} \ln A_{i,t-1} + \epsilon_t$; $\epsilon_t \thicksim \mathcal{N}(0, \sigma_{A})$ and affects the production, given by $Y_{it} = A_{it} K_{it}^{1-\alpha} N_{it}^\alpha$, of a sector. $\alpha = 2/3$, $\rho_{A} = 0.97$, $\sigma_{A} = 0.1$, and $A_{i0} = 1$. The agents are as in \S\ref{section_global}, being statically heterogeneous in propensity to work, $\theta$ and employer $e$, and dynamically heterogeneous in investment, $k_t$; however now have visibility of all prices, not just wages. The global model that couples the agents is the same RBC model described in \S \ref{section_epi} and Appendix I, but \textit{without} infections. As the agents' internal states are no longer stochastic, a smaller number ($J=10$) of agents can be used.

\begin{figure}[ht]
\vskip 0.2in
\begin{center}
\centerline{\includegraphics[width=\columnwidth]{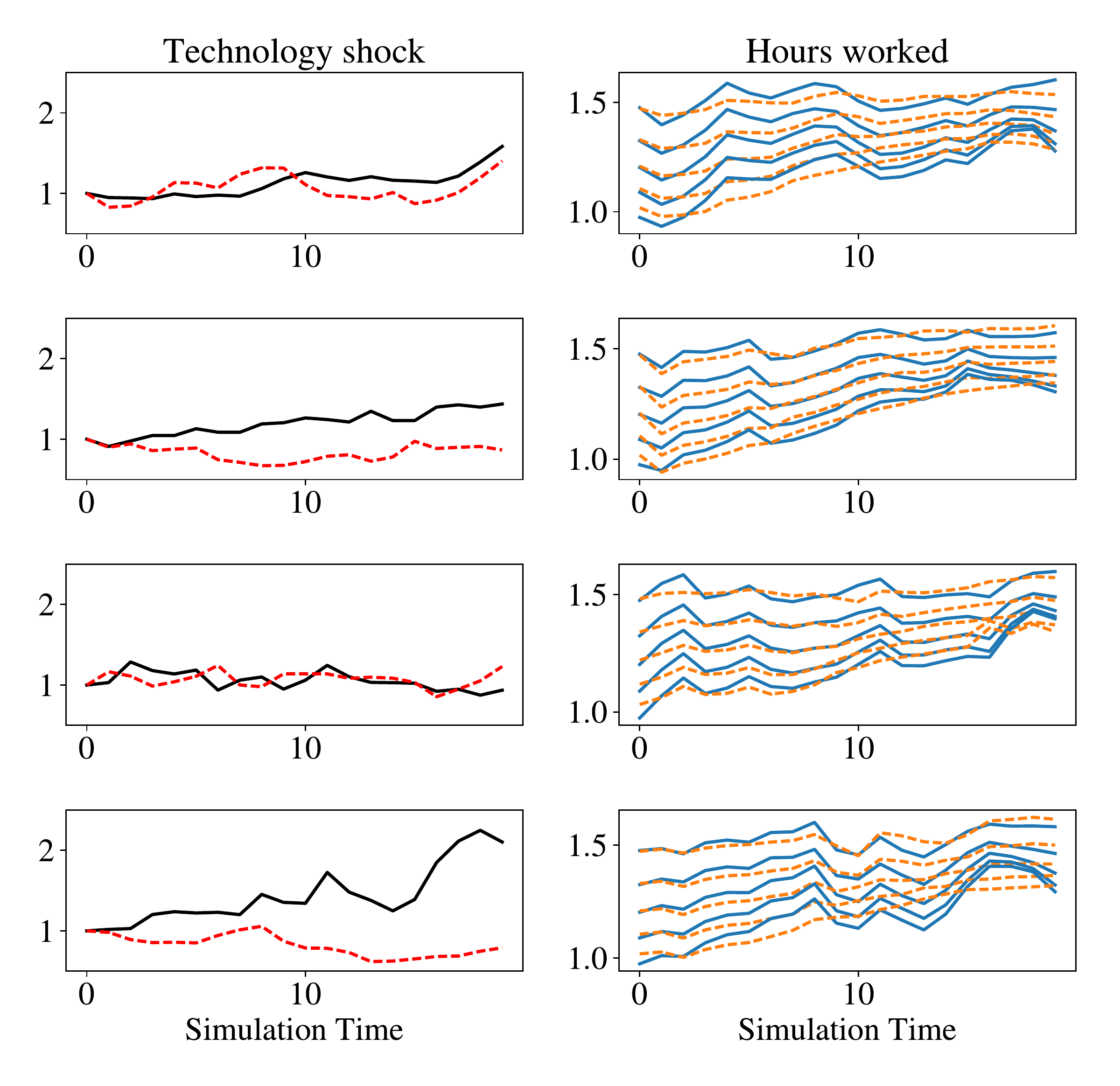}}
\caption{Each row represents one of 4 randomly chosen simulations. \emph{Left column} The technology shock for sectors $0$ (black, solid) and $1$ (red, dashed). \emph{Right column} Paths of hours worked for agents who can (orange, dashed) and cannot (blue, solid) observe realisation of prices when choosing their action; all agents work for sector $0$ and have $\theta = \{0.76,0.88,1.0,1.12,1.24\}$ from top to bottom. }
\label{fig-hours-savings}
\end{center}
\vskip -0.2in
\end{figure}
As in \S \ref{section_global}, we compare two types of agents, one that `sees' realisations of the prices and another than does not. Since differing values of the technology shock move the general equilibrium, changing the prices, observing them gives the agent information about the state of the underlying stochastic process. Figure \ref{fig-hours-savings} shows the paths of hours worked, $n_j$, for $5$ agents with a range of values of $\theta$, all of whom work in the same sector given by $i=0$. As expected, the paths of agents who can observe realisations of prices have smaller fluctuations, which is also true of the paths of other quantities. Averaging over $256$ runs, the mean unsigned curvature of the paths drops from $\bar{\kappa}_{non-obs} = 0.44$ to $\bar{\kappa}_{obs} = 0.27$; this fall is also true of the curves in the figure where $\bar{\kappa}_{non-obs} = 0.42$ and $\bar{\kappa}_{obs} = 0.19$. This difference is because agents who can observe realisations are better able to adjust their behaviour to the current and (since it is an autoregressive process) future values of the technology shock.

\subsubsection{Specification}

The specification of the neural network and learning remains unchanged from the previous section. Calculation of the multiple histories is parallelised; we use $8$ threads. For each simulation epoch $E$, $8(4+E)$ histories are found, followed by $20$ RL training epochs. In total, there are $12$ simulation epochs and a total of ${\approx}1000$ histories and $240$ RL training epochs, each of which samples from the most recent $50\%$ of the histories. The number of histories and epochs is informed by the convergence properties from the previous section. A total of ${\approx}100,000$ experiences are recorded during the whole simulation which takes ${\approx}6$ hours.

\section{Discussion and Conclusion}

This work shows that reinforcement learning can be used to solve a wide range of important macroeconomic rational expectations models in a way that is simple, flexible, and computationally tractable. Furthermore, these methods can be immediately applied to previously intractable problems with multiple degrees of discrete heterogeneity and stochasticity. Being highly relevant to real world phenomena, such as climate change and disease transmission, these capabilities are of great value to policymakers and can be developed into serious tools to aid decision making in complex scenarios.

Finally, by linking to reinforcement learning, this work provides the potential to apply its extensive toolkit of techniques, many of which have direct relevance to economic questions: examples include accessing larger state and action spaces \citep[e.g.][]{lillicrap2015continuous}, including bounded rationality, or applying inverse reinforcement learning to deduce agents' objectives and rewards from observed micro- and macro-economic behaviours. Additionally, we can harness improvements in implementation such as GPU/TPU acceleration \citep{NEURIPS2019_9015} and distributed computing \citep{mnih2016asynchronous}.
\section*{Acknowledgements}

We would like to thank Federico Di Pace for useful discussions.
\goodbreak

\bibliography{bib_op}
\bibliographystyle{icml2020}

\section*{Appendix I : The Real Business Cycle model}

We use a standard real business cycle model, however we adopt notation and variables common in reinforcement learning, in particular, an emphasis on state variables (capital) rather than action variables (consumption, hours worked), and the inclusion of an action-dependent expectation. A baseline RBC model would use Equations \ref{bc}, \ref{rbc1}, \ref{rbc2}, \ref{pf}, \ref{dn}, \ref{dk}, and \ref{shock}. We use Equations \ref{bc}, \ref{ee}, \ref{ff}, \ref{pf}, \ref{dn}, \ref{dk}, and \ref{shock}, but note that \ref{ee} and \ref{rbc1}, and \ref{ff} and \ref{rbc2}, are the same up to algebraic manipulation, expressing the future behaviour in terms of $U(k)$ and $U'(k)$, functions of the state, rather than in terms of consumption or other actions as is usually seen.

We work in real quantities, using $p_{i=0}=1$ as the num\'eraire and other quantities, including $p_{i\neq0}\neq 1$ defined relative to this.

\subsection*{Notation}
$j$ is an index that runs over the $J$ consumer-workers. $i$ runs over the $I$ consumption goods, each of which is produced by a different sector/firm. 

For consumers, $n_j$ is hours worked, $c_{ji}$ is consumption, $k_j$ is capital held by the consumer, $v_j$ is their investment in capital in that timestep, $\theta_j>0$ is the weight given to hours in the utility function. $E_i$ denotes the set of agents employed at firm $i$, $e(j)$ is the index, $i$, of the employer of $j$. 

For firms, $N_i$ is the number of hours worked at the firm, $K_i$ is its capital, $A_i$ is the firm's technology, $Y_i$ is the production function, $C_i$ is the consumption of the firm's goods.

$r$ is the real interest rate, $w_i$ are wages, and $p_i$ are the prices of goods.

\subsection*{Consumers}
For consumers, the time-$t$ utility is $$
    u_j(\{c_{ji}\}, n_j; \theta_j) = \sum_i \ln{c_{ji}} - \frac{1}{2}\theta_j n^2_j
$$
and their total utility from time $t$ onward is 
\begin{multline}
U_{j,t,S}(k_{t,j}) =\\ \max_{c_{ji},n_j} \left\{u_{j}(c_{ji}, n_j) + \beta \sum_{S'} P_{S\rightarrow S'}(c_{ji}, n_j)U_{j,t+1,S'}(k_{t+1,j})\right\}
\end{multline}
Their budget constraint is,
\begin{equation}
      w_{e(j)} n_j + r k_{j,t} = \sum_i p_i c_{ji} + v_j \;\;\;\; \forall j \label{bc}
\end{equation}
$$
      k_{j, t+1} = k_{j,t} + v_j
$$

Let $U_j' = \partial_{k_{t+1,j}}U(k_{t+1,j})$; expectations are over a distribution of probabilities $P_{S\rightarrow S'}$, and so $\expect \partial_{c_{ji}} \ln P_j = \sum_{S\rightarrow S'} \partial_{c_{ji}}P_j$. Consumers take prices ($p_i$, $w_i$ and $r$) as given and use their first order conditions
\begin{equation}
    0 = c_{ji}^{-1} + \beta \expect \left( U_j \partial_{c_{ji}} \ln P_j- \beta p_i U_j'\right) \label{ee}
\end{equation}
\begin{equation}
    0 = -\theta_j n_j +  w_{e(j)}\beta \expect U_j' \label{ff}
\end{equation}
to find $c_{ji}$ and $n_j$; this is solved iteratively since $U$ is a function of $k_{t+1}$ and thus $c_{ji}$ and $n_j$. In the reinforcement learning training we add an additional reward for individually achieving a no-Ponzi condition at the final time-step.

If the probabilities were independent of the action, $\partial_{c_{ji}}P_j = 0$, as would be the case in a standard RBC model, then that term can be removed and the $\expect U_j'$ eliminated obtaining 
\begin{equation}
n_j = \frac{w_{e(j)}}{ \theta c_{ji} p_i}\;\; \forall i \label{rbc1}
\end{equation}
A small amount of work, with care taken as to the maximisation over $a$ in the definition of the utility, shows that $\expect U_j' = (1+r_{t+1})(p_{i,t+1}c_{ji, t+1})^{-1}$, and so Equation \ref{ee} reduces to the Euler equation 
\begin{equation}
c_{ji}^{-1} = \beta \expect \left[\frac{p_{i}}{p_{i,t+1}} (1+r_{t+1}) c_{ji, t+1}^{-1}\right] \label{rbc2}
\end{equation}
where the $p_i$ remains due to the multiple goods with $p_{i\neq0}\neq1$.

\subsection*{Firms}
Firms are profit maximising with production function
\begin{equation}
    Y_i = A_i K_i^{1-\alpha}N_i^{\alpha}\label{pf}
\end{equation}
Since profit $\Pi_i =  p_i Y_i - w_i N_i - (r + \delta)K_i$, then taking prices ($p_i$, $w_i$, $r$) as given 
\begin{equation}
    \partial_{N_i} \Pi_i = 0 \;\;\;\;\;\; p_i \alpha Y_i - w_i N_i = 0 \;\;\; \forall i \label{dn}
\end{equation}
\begin{equation}
    \partial_{K_i} \Pi_i = 0\;\;\;\;\;\; p_i (1-\alpha) Y_i - (r + \delta) K_i = 0\;\;\; \forall i \label{dk}
\end{equation}
and therefore $\Pi_i = 0$.
\begin{equation}
      K_{i, t+1} = K_{i,t}(1-\delta) + (Y_i - C_i)
\end{equation}
We split $K_i = K_e + \hat{K}_i$ where $K_e$ is an endowment and $\hat{K}_i$ is provided by investment from the consumers.
\begin{equation}
      \tilde{K}_{i, t+1} = \tilde{K}_{i,t}(1-\delta) + (Y_i - C_i) - (r+\delta)K_e
\end{equation}

\subsection*{The technology shock}
The technology shock is log-autoregressive
\begin{equation}
    \ln A_{it} = \rho_{A} \ln A_{i,t-1} + \epsilon_t\;\;\;\;\epsilon_t \thicksim \mathcal{N}(0, \sigma_{A}) \label{shock}
\end{equation}

\subsection*{Market clearing conditions}
Wages are set by market clearing for hours worked
\begin{equation}
    N_i = \sum_{j \in E_i}g_j n_j \;\;\;\;\; \forall i\label{mchw}
\end{equation}

and the real interest rate is set by market clearing for capital
\begin{equation}
    \sum_i \tilde{K}_i = \sum_{j}g_j k_j
\end{equation}
where $g_j$ is the weight of each agent, with $\sum_j g_j = 1$.

\end{document}